\documentclass[aps,prd,twocolumn,showpacs,preprintnumbers]{revtex4}

\usepackage[dvips]{graphicx}
\usepackage{latexsym}
\usepackage{graphicx,graphics,epsfig,epstopdf,subfigure,color,psfrag}
\usepackage[english]{babel}
\usepackage[latin1] {inputenc}
\usepackage{amsmath,amsfonts,amssymb}
\usepackage{bm}
\usepackage{color}

\newcommand \be {\begin{equation}}
\newcommand \bea {\begin{eqnarray}}
\newcommand \ee {\end{equation}}
\newcommand \eea {\end{eqnarray}}
\newcommand \bed {\begin{displaymath}}
\newcommand \eed {\end{displaymath}}

\newcommand{\bit}{\begin{itemize}}
\newcommand{\eit}{\end{itemize}}

\newcommand{\bgar}{\begin{eqnarray}}
\newcommand{\enar}{\end{eqnarray}}

\begin{document}
\preprint{CERN-PH-TH/2013-281}
\title{Nuclear Structure Functions at Low-$x$ in a Holographic Approach}
\author{L. Agozzino$^{a,b}$, P. Castorina$^{a,b,c}$, P. Colangelo$^{d}$}
\affiliation{$^{a}$Dipartimento di Fisica, Universit\'a  di Catania, via S. Sofia 62, I-95125 Catania, Italy\\
$^{b}$INFN, Sezione di Catania, via S. Sofia 62, I-95125 Catania, Italy\\
$^{c}$PH Department, TH unit, CERN, CH-1211 Geneva 23, Switzerland\\
$^{d}$INFN, Sezione di Bari, via Orabona 4, I-70126 Bari, Italy}

\begin{abstract}
Nuclear effects in deep inelastic scattering  at low$-x$ are phenomenologically described  changing the typical dynamical and/or kinematical 
scales  characterizing the free nucleon case. In a holographic approach, this rescaling is an analytical property of the computed structure function $F_2(x,Q^2)$. This function  is given by the sum of a conformal term and of a contribution due to quark confinement, depending on IR hard-wall parameter $z_0$ and on the  mean square distances, related to a parameter $Q^\prime$, among quarks and gluons in the target. The holographic structure function per nucleon in a nucleus $A$ is evaluated showing that  a rescaling of the typical nucleon size, $z_0$ and $Q^\prime$, due to nuclear binding, can be reabsorbed in a  $Q^2$-rescaling scheme.  The difference between neutron  and proton structure functions and the effects of the longitudinal structure functions  can also be taken into account. The obtained theoretical results favourably compare with the experimental data.

\end{abstract}
\pacs{11.25.Tq, 11.10.Kk, 12.38.Lg, 24.85.+p}
\maketitle

\section{Introduction}

Deep inelastic scattering (DIS) experiments of charged leptons off nuclei have shown that the structure functions of nucleons bound in nuclei differ from the structure functions of free, isolated nucleons. Although in some cases a deviation could be expected considering, e.g.,   the Fermi motion of nucleons in nuclei, in general the interpretation and the predictions of  the nuclear modifications have  presented considerable difficulties and,  not surprisingly,  the measurements have generated an  intense theoretical and phenomenological activity \cite{arneodo,Frankfurt:2011cs}. 

Nuclear effects can be described  comparing the structure functions of the nuclear target, normalized to the number of  nucleons,  to the free nucleon ones. For electroproduction,
if $F_2^D$  is the structure function of the deuterium $D$ and $F_2^A$  the structure function per nucleon of  the nucleus $A$, 
 the ratio $R_A= F_2^A(x,Q^2)/F_2^D(x,Q^2)$ is measured for various values of  the Bjorken variable $x$ and the squared momentum transfered $Q^2$. Nuclear modifications are observed to depend on  $x$. For  $x \le 0.1$  the ratio  $R_A$ is found  $R_A < 1$: this is the so-called  shadowing region.   In the range $0.1 < x< 0.25$ there is the anti-shadowing,  with $R_A > 1$. 
For large $x$  the so-called EMC effect appears: again a decreasing behaviour.

There are different approaches aimed at  interpreting such observations. A few of them make use, both for the EMC and the shadowing effect, of the idea that the nuclear modifications are mainly due to the  change of the effective
mean square distances among quarks and gluons in a nuclear environment with respect to  free nucleons \cite{arneodo}. Such a geometric modification can be accounted for 
by a rescaling of the kinematical variables, $x$ or $Q^2$, in the structure functions of  a free nucleon. This is the case, for example, 
of the so-called  $x$-rescaling model, where the EMC effect is described by rescaling  the Bjorken $x$ variable  in the free nucleon $F_2^D$ \cite{aku}:
\be
F_2^A(x,Q^2)=F_2^D(x/\hat z,Q^2) \,\,\, .
\ee
The factor $\hat z$ is  defined as $\hat z \simeq 1- \epsilon/M$, in terms of     the proton mass $M$ and of  the energy $\epsilon$ necessary to emit a nucleon from a nucleus. A difficulty of this model is that the values of the energy  $\epsilon$  to  fit the large-$x$ data  exceed the calculations of the nuclear binding  \cite{arneodo}.

The  $Q^2$-rescaling model of the EMC effect is based  on the relation \cite{close,close2}
\be
F_2^A(x,Q^2)=F_2^D(x, \chi_A Q^2) \,\,\, , 
\ee
indicating that the effective $Q^2$ for a bound nucleon is different from the free one. Such a  dynamical property is related to the modification of the quark confinement scale in the nucleus \cite{close,close2}:
quarks and gluons are no longer confined to specific nucleons, but  spread over distances larger than the  free nucleon size. By studying the moments of the structure function,  starting from a $Q^2$ region where the valence picture is a good approximation,  one can show that in QCD, for large $Q^2$, the  change of scale is related to the strong coupling constant $\alpha_s$.  It is worth remarking that the $x$- and $Q^2$-rescaling models, although different in their  assumptions,  can be  related \cite{close4}.

A different nonperturbative approach considers that the low-$x$ region is governed by the Pomeron exchange \cite{castorina}. In a nuclear environment, the nucleon overlap produces a suppression of  the effective  quark-Pomeron coupling. Indeed, although quarks and gluons are no longer confined to specific nucleons  and spread over distances larger than the  free nucleon size, the average spatial separation between the quarks before color neutralization decreases,  and this reduces the Pomeron coupling  which 
 is   related to  such a typical size \cite{pov}.
 
The idea that the description of the nuclear modifications  requires to evaluate the  change of the free nucleon wave function induced by the  nuclear binding can find  a support in an
analysis based on the holographic approach.
The AdS/CFT, or gauge/gravity correspondence principle \cite{adscft} is important to access the nonperturbative sector of gauge theories,  and can be used  to study  features of QCD  \cite{finite}. The method has been applied to DIS at strong coupling \cite{Polchinski,other-dis}.  In particular, at low-$x$
the nucleon structure function  $F_2^N(x,Q^2)$ has been computed  in Ref. \cite{Brower:2010wf},  and has been represented as a conformal contribution and  an additional term accounting for  quark confinement. Both contributions involve the holographic nucleon wave function:
since the confinement dynamics  determines the modification of the structure functions of a nucleon in nuclei,  the holographic baryon wave function in  nuclei affects the
nuclear structure functions.
 Following this viewpoint, in the study \cite{noi} we  attempted a description of  shadowing in  a gauge/gravity framework,  using in the low-$x$ region the AdS/CFT strong coupling BPST Pomeron kernel  computed in  \cite{Brower:2006ea}. The holographic free nucleon wave function is assumed to be peaked at  a distance $1/Q^\prime$  close to the boundary $z_0$.
In the  description of the nuclear binding effects,  the  wave function of the bound nucleon must involve a different effective distance $1/Q^\prime_A$ and a new confinement boundary $z_0^A$.  Studying the scaling properties of the  holographic expression for $F_2$ under the replacement $Q^\prime \rightarrow  Q^\prime_A $ and $z_0 \rightarrow z_0^A$, nuclear effects turn out to be described  by a  rescaling of the confinement parameters, with a remarkable agreement with measurements.

Here,  we  discuss this idea in more details,  including the difference between proton and neutron structure functions,  analyzing a few  approximations adopted in Ref.\cite{noi},  considering  the $x$-rescaling scheme,  carrying out a more complete comparison with the  experimental data, evaluating the effects of the longitudinal structure function. 
The paper is organized as follows: in   Sec.~\ref{sec:proton} we review the low-$x$ behavior of the proton structure functions in a holographic approach, and discuss the neutron-proton difference.  Sec.~\ref{sec:nuclearsf} contains the model for the nuclear modifications of the structure functions, which is compared with  data in Sec.~\ref{sec:nuclei}.   In Sec.~\ref{sec:FL} we discuss the longitudinal structure function in nuclei, 
and in Sec.~\ref{sec:conclusions} we present our  conclusions.

\section{Holographic  Proton Structure Functions}\label{sec:proton}

The AdS/CFT calculation of  DIS at low-$x$ on a proton was first considered by  Polchinski and Strassler  in  \cite{Polchinski}.  After this seminal proposal, several calculations have been carried out
in various holographic frameworks \cite{other-dis}. In particular,  in
\cite{Brower:2010wf}   the nucleon structure function $F_2$ was computed analyzing the virtual $\gamma^* p$ total cross section, and  two contributions were obtained, a term for conformal gauge theories and an additional term accounting for confinement. A slice  of the dual AdS space  was used to break the  conformal invariance.
As shown in  \cite{noi}, this result  can be used to analyze nuclear effects on $F_2$.

The definition in QCD of the structure functions $F_1(x,Q^2)$ and $F_2(x,Q^2)$ of a a hadron of momentum $P$ and charge $\cal Q$ is based on  the matrix element of two electromagnetic currents 
\be
T^{\mu \nu} \equiv i \int d^4y e^{i q \cdot y}  \langle P {\cal Q} | T \left(J^\mu(y) J^\nu(0) \right) | P {\cal Q} \rangle \,\,\,  , \label{eq:Tmunu}
\ee
which can be written as
\bea
T^{\mu \nu} &=& F_1(x,Q^2)\left(\eta^{\mu \nu}-\frac{q^\mu q^\nu}{q^2}\right) \nonumber \\
&+& \frac{2 x}{q^2}F_2(x,Q^2)\left(P^\mu+\frac{q^\mu}{2x}\right)\left(P^\nu+\frac{q^\nu}{2x}\right) 
\,\,\, .\label{eq:Tmunu12}
\eea
 $\mu,\nu$ are four-dimensional indices, $\eta^{\mu \nu}$  the Minkowski metric,  the Bjorken variable $x$ is  $\displaystyle x=\frac{Q^2}{2 P \cdot q}$, with  $Q^2=-q^2$.

The AdS/CFT calculation involves $R$-currents in (\ref{eq:Tmunu}), and the couplings 
\be
g_s=\frac{g_{YM}^2}{4 \pi}=\alpha_{YM}=\frac{\lambda}{4 \pi N_C},  \,\,\,\,  R=\alpha^{\prime \frac{1}{2}} \lambda^{\frac{1}{4}} \,\,\, . \label{couplings}
\ee
$g_{YM}$ is the Yang-Mills coupling constant, $N_C$ the number of colors, in the regime 
 $g_s<<1$ and $\lambda >>1$.   $R$ is the $AdS$ radius.
 
The dual string calculation of the matrix element (\ref{eq:Tmunu}),  or of its imaginary part  appearing in DIS processes, describes the photon-hadron scattering  $\gamma^*p \to \gamma^* p \equiv 1,2 \to 3,4$ as occurring in the AdS space.
Various quantities are needed, starting from the states dual to the initial-final hadron $p$. For protons,  these states are represented by normalizable wave functions $\phi^p(z)$,  in principle obtained  from a suitable equation of motion, with some dependence on the holographic coordinate $z$. 
For  the calculation of the matrix element (\ref{eq:Tmunu})  the transition  function is required:
\be 
P_{24}(z)=\sqrt{-g} \, \left(\frac{z}{R}\right)^2 \, \phi^p(z) \phi^p(z) \,\,\, .
\ee 
  
The current that couples to the hadrons in the matrix element  (\ref{eq:Tmunu}) excites non-normalizable modes of the gauge fields $\cal A$, which in the bulk obey Maxwell's equations.  In the Lorentz gauge and for $R=1$  there are the solutions:  ${\cal A}_\mu(y,z) =  n_\mu (Qz) K_1(Qz) e^{i q\cdot y}$ and ${\cal A}_z(y,z) = i (q \cdot n) (Qz) K_0(Qz) e^{i q\cdot y}$,
 given in terms of Bessel functions $K_1$ and $K_0$ and of  the  polarization vector $n_\mu$.
The calculation of the structure function $F_2$ in (\ref{Tmunu12},\ref{eq:Tmunu})  requires  the transition function
\be
P_{13}(z,Q^2) = \frac{1}{z} (Qz)^2 \left[K_0^2(Qz) + K_1^2(Qz)\right],  \label{P13}
\ee
with $Q=\sqrt{Q^2}$, while 
\be
P_{13}(z,Q^2) = \frac{1}{z} (Qz)^2  K_1^2(Qz)  \label{P13prime}
\ee
is needed for $2xF_1$.
From now on, we focus on $F_2$: the effect of the nuclear modification on the longitudinal structure function $F_L = F_2 - 2xF_1$ will be discussed in Sec. \ref{sec:FL}.

Finally, the scattering kernel is needed. Expressing it in terms of a Pomeron Regge pole contribution  \cite{Brower:2006ea},   at low-$x$ the structure function $F_2$ can be written as an eikonal sum \cite{Brower:2010wf}:
\bea
F_2^p(x,Q^2) &=& \frac{Q^2}{2 \pi^2}
 \int d^2 b \int dz dz^\prime P_{13}(z,Q^2) P_{24}(z') \nonumber \\ &\times &{\rm Re} \left( 1- e^{i \chi (s,b,z,z^\prime)}\right) \,\,\, . \label{F2N}
\eea
$s$  is the center-of-mass energy squared of the $\gamma^*$-target system and $b$  the impact parameter. The derivation  of 
the eikonal $\chi$   for conformal theories and including conformal breaking effects is in Refs.~\cite{Brower:2006ea,Brower:2010wf}.

\subsection{Conformal term}

An  expression of the  proton structure function $F_2^p $ in the conformal case, derived from   Eq.(\ref{F2N}), as been worked out in Ref.~\cite{Brower:2010wf}:
\bea
F_{2cf}^p(x,Q^2) &=& \frac{g_0^2 \rho^{3/2}}{32 \pi^{5/2}}   \int dz dz^\prime \frac{z z^\prime Q^2}{\tau^{1/2}} P_{13}(z,Q^2 ) P_{24}(z^\prime) \nonumber \\
&\times& e^{(1-\rho) \tau} \exp{[\Phi(z,z{'},\tau)]} \,\,\, . \label{F2Nconf}
\eea
$g_0^2$ is a parameter and $x \simeq Q^2/s$;  $\rho$ is defined in terms of the 't Hooft coupling in (\ref{couplings}),     $\displaystyle \rho=2/\sqrt{\lambda}$.
The function $\tau$,  defined as  $\tau = \log{(\rho z z^\prime s/2)}$,   is a conformal invariant.
$\Phi$ is the BPTS  Pomeron kernel integrated in impact parameter   \cite{Brower:2006ea}:
\be
\Phi(z,z^\prime,\tau) = - \frac{(\log{z} - \log{z^\prime})^2}{ \rho \tau }\,\,\, .
\ee

\begin{figure}[t]
\includegraphics[width = 0.4\textwidth]{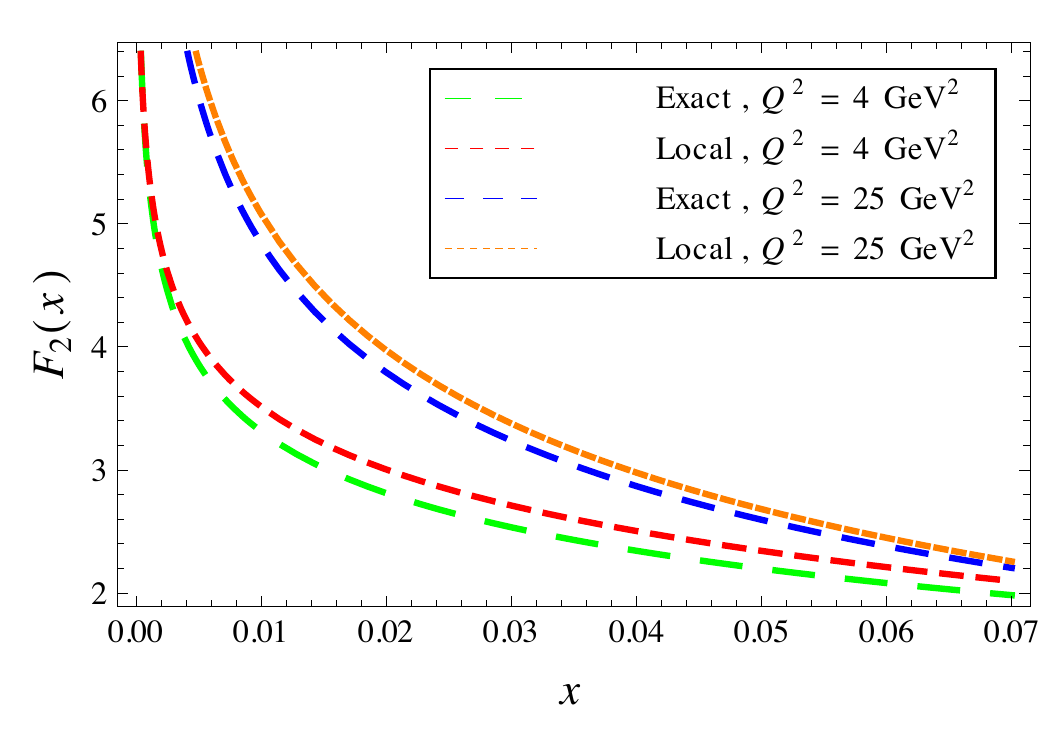}
\caption{Comparison between  $F_{2cf}^p$ in Eq.(\ref{F2Nconf}),  obtained  using  Eq.(\ref{P13}) (labeled as Exact) and  the approximation  Eq.(\ref{localP13}) (labeled as Local),  for  $Q^2=4$ and $25$ GeV$^2$.}\label{fig:F2local}
\end{figure}

Eqs.(\ref{F2N}) and (\ref{F2Nconf}) involve the transition functions $P_{24}$ and $P_{13}$.
 The proton  wave function in the bulk $\phi^p(z)$,  needed in $P_{24}$,   should be determined by an explicit  holographic model for the baryon.
An approximation has been used in Ref.~\cite{Brower:2010wf},  assuming that  $\phi^p(z)$ is  peaked close to the infrared boundary $z_0$, with $1/Q^\prime \leq z_0$ and $Q^\prime$ of the order of nucleon mass, giving:
\be
P_{24}(z^\prime) \simeq \delta\left(z^\prime - \frac{1}{Q^\prime}\right) \,\,\, . \label{localP24}
\ee
Moreover, also $P_{13}$ can be replaced by a local expression
\be
P_{13}(z, Q^2) \simeq C \delta\left(z - \frac{1}{Q}\right) \,\,\, , \label{localP13}
\ee
with $C \simeq 1$ \cite{Brower:2010wf}.  This  is justified by the shape of  the function
$P_{13}$ in Eq.(\ref{F2Nconf}),  which  is peaked for $z \simeq 1/Q$. 
In Fig.~\ref{fig:F2local} we depict $F_{2cf}^p$ obtained  using the exact expression in Eq.(\ref{P13}) and the local approximation  Eq.(\ref{localP13})  for two values of the squared transfered momentum:
the relative difference between the  two expressions  is within few per cent for  $x<0.07$.

The resulting  $F_{2cf}^p$ reads \cite{Brower:2010wf}:
\be
F_{2cf}^p(x,Q^2,Q^\prime) = \frac{g_0^2 \rho^{3/2}}{32 \pi^{5/2}} \frac{Q}{Q^\prime} \frac{e^{(1-\rho) \tau} }{\tau^{1/2}} 
e^{-\left[\log^2{(Q/Q^\prime)}/\rho \tau\right]} \,\, . \label{F2Nconf1}
\ee

\subsection{Confinement term}

The  expression for the  proton structure function $F_{2cf}^p$,  based on the conformal BPST Pomeron, does not  fit the HERA data in the low-$Q^2$ range, where confinement is the main dynamical mechanisms \cite{Brower:2010wf}.  Confinement can be described in the holographic approach  including an infrared boundary  $z_0$ on the $z$ bulk coordinate, a so-called 
 hard-wall holographic model of QCD.  This confinement scale can be related to  $\Lambda_{QCD}$.  
The eikonal is modified and a non-conformal contribution to $F_2^p$ should be considered, which  reads  for a single Pomeron \cite{Brower:2010wf}:
\bea
F_{2ct}^p(x,Q^2,z_0) &=& \frac{g_0^2 \rho^{3/2}}{32 \pi^{5/2}} \int dz dz^\prime \frac{z z^{\prime}Q^2}{\tau^{1/2}} P_{13}(z,Q^2 ) P_{24}(z^\prime) \nonumber \\
&\times& e^{(1-\rho) \tau} \,\, e^{- \frac{\log^2{\left(z z^\prime/z_0^2\right)}}{ \rho \tau}}\,\, G(z,z^\prime,\tau) . \nonumber \\ \label{F2Nnew}
\eea
The  $z_0$ dependence  is shown explicitly. The function $G(z,z^\prime,\tau)$ is 
\be
G(z,z^\prime,\tau) = 1- 2 \sqrt{\rho \pi \tau} e^{\eta^2} erfc(\eta) \,\,\, , \label{G2}
\ee
with
\be
\eta =\frac{-\log{\left(zz^\prime/z_0^2\right)} + \rho \tau}{ \sqrt{\rho \tau} } \,\,\, .\label{eta}
\ee
Adopting  the  approximation (\ref{localP24}) and (\ref{localP13}),  Eq.(\ref{F2Nnew}) reduces to
\bea
F_{2ct}^p(x,Q^2,Q^\prime,Q_0^2) &=& \frac{g_0^2 \rho^{3/2}}{32 \pi^{5/2}} \frac{Q}{Q^\prime} \frac{e^{(1-\rho) \tau} }{\tau^{1/2}} \nonumber \\  
&\times& e^{- \frac{\log^2{\left(Q_0^2/\left(Q Q^\prime\right)\right)}}{ \rho \tau}} G\left(\frac{1}{Q},\frac{1}{Q^\prime},\tau\right) \,\,\, , \nonumber \\
\label{F2NQ0}
\eea
with $Q_0=1/z_0$ \cite{Brower:2010wf}.

The proton structure function $F_2^p$ results from the sum of the conformal and confinement contribution,
\be 
F_{2}^p(x,Q^2) = F_{2cl}^p(x,Q^2,Q^\prime) + F_{2ct}^p(x, Q^2,Q^\prime,Q_0^2)\,\,\, ,  \label{F2p}
\ee
and can successfully be compared with proton DIS data  \cite{Brower:2010wf}.

\begin{figure}[h]
\includegraphics[width = 0.4\textwidth]{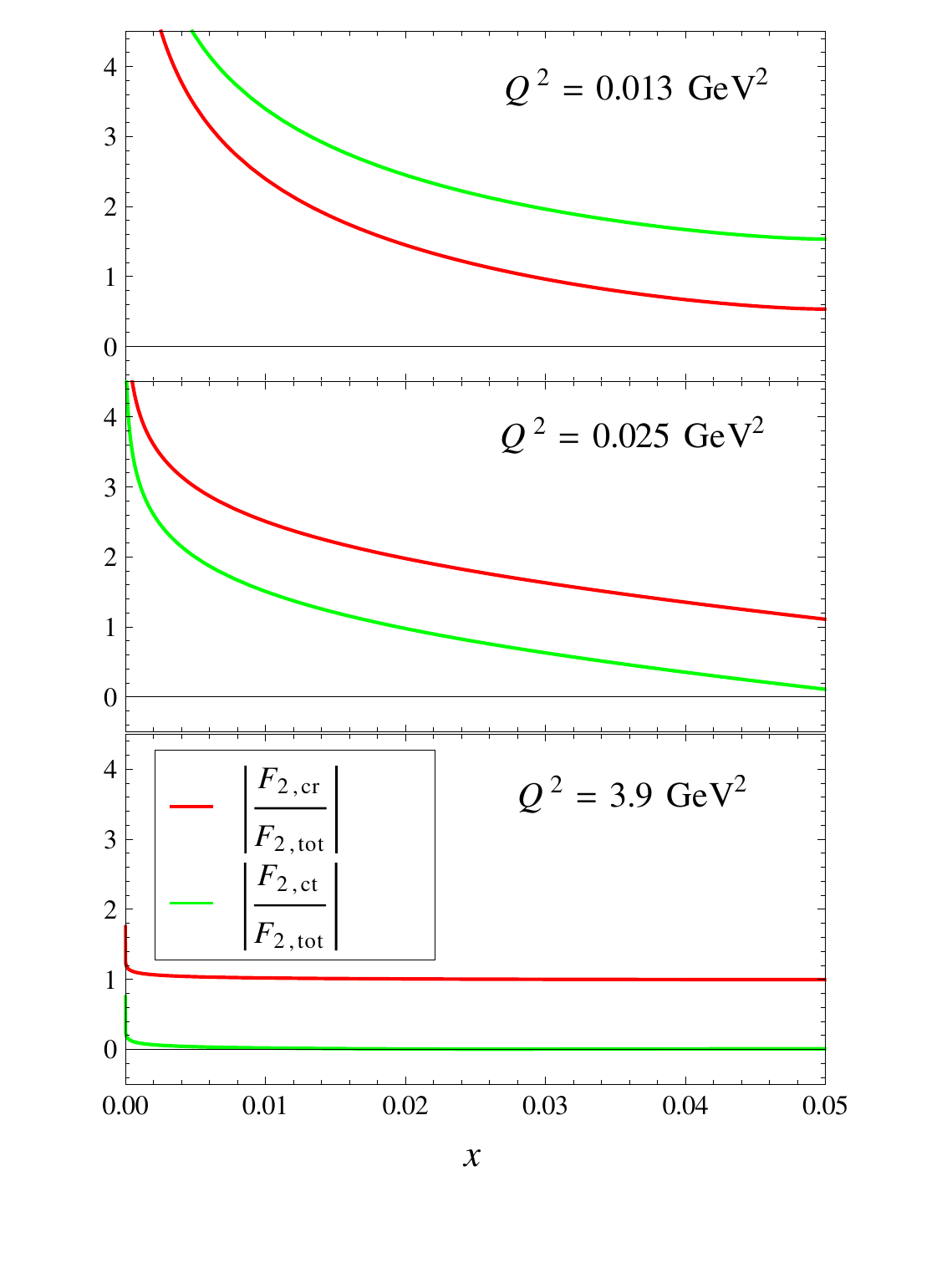}
\caption{Comparison between  conformal and confinement contributions to  $F_2^p(x,Q^2)$  at low-$x$, for three  values of $Q^2$. The red (dark) lines correspond to the absolute value  
$|F_{2cf}^p/F_2^p|$ of the ratio of the conformal term $F_{2cf}^p$ in Eq.(\ref{F2p})  over the full structure function;  the  green (light)  lines correspond to   ratio 
$|F_{2ct}^p/F_2^p|$ of the confinement contribution $F_{2ct}^p$  to the full structure function.   For the lowest value of $Q^2$ the confinement term dominates.} \label{figcfct}
\end{figure}

It is interesting to analyze the relative weight of the conformal and confinement  contributions to $F_2^p$ at low-$x$.  In Fig.~\ref{figcfct}  three values of $Q^2$ are considered:  at  $Q^2\simeq 4$ GeV$^2$ the structure function is essentially determined by the conformal term.  On the other hand, the confinement term is  the main contribution  at very low $Q^2$  for all the considered values of the Bjorken-$x$. 

\subsection{Accounting for isospin effects: neutron structure function}

Isospin effects play an important role in  detailed analyses of nuclear structure functions (normalized to the total number of nucleons). These effects represent the difference between the proton and neutron structure function.
In the holographic model, the difference can be implemented in a rather simple way replacing the scales $Q_{0}$ and $Q^\prime$ for the proton
with corresponding scales $Q_{0n}$ and $Q^\prime_n$ for the neutron.  Therefore, the neutron structure function $F_2^n$ can be represented by the expression 
\be
F_{2}^n = F_{2cl}^p(x,Q^2,Q'_n) + F_{2ct}^p(x, Q^2,Q'_n,Q_{0n}^2)\,\,\,, \label{neut}
\ee  
with $Q_{0n} \simeq Q_0$, since  proton and neutron have a  similar color confinement scale. 

The experimental information on the neutron structure function comes from DIS on a deuterium target;  therefore,  the comparison of the expression (\ref{neut}) with data  requires implementing the nuclear effects discussed in the next section.  Here we anticipate  the  proposal to describe the isospin difference in the holographic formula mainly  through the parameter $Q^\prime_n$.

\section{Nuclear structure functions in holographic framework}\label{sec:nuclearsf}

 In the Introduction we have mentioned  that a physical description of the EMC and of the shadowing effects can be obtained considering an  effective modification of the dynamical length/momentum scales in deep inelastic scattering processes on a nuclear target  with respect to a free nucleon.
It is remarkable that such a rescaling,  in particular the $Q^2$ rescaling,  is a property of the analitic expression of the holographic structure function, not only in the conformal term but also in the term taking  the confinement dynamics  into account. 

Let us focus on the conformal contribution  (\ref{F2Nconf1}) to
$F_2^N$ ($N=$ nucleon, neglecting for the moment the proton-neutron difference), which  depends on the ratio $Q/Q^\prime$.  The description of the  modification of the structure function (per nucleon) $F_2^A$ in the nucleus $A$, using the rescaling
\be
Q^\prime_A = \lambda_A  Q^\prime, \label{Qresc}
\ee
  corresponds to the rescaling $Q^2 \rightarrow Q^2/\lambda_A^2$. 
In (\ref{Qresc})  $Q^\prime_A$  is identified with the typical scale of the wave function of the bound nucleon. Consequently, 
 one has
\be
F_{2cf}^A(x,Q^2) = F_{2cf}^N\left(x, \frac{Q^2}{\lambda_A^2}, Q'\right) \,\,\, ,
\ee
and the $Q^2$-rescaling at low-$x$ naturally arises  in the conformal contribution to the holographic expression of $F_2$.  

In the confinement term in  Eqs.(\ref{F2Nnew}), (\ref{G2}) and (\ref{eta})   a nontrivial $Q^2$ behavior appears  in the log-factors and in  $\eta$,  due to the  infrared scale $Q_0$.
The  rescaling $Q^\prime_A = \lambda_A  Q^\prime$ can be reabsorbed in the  $Q^2$ rescaling,  $Q^2 \rightarrow Q^2/\lambda_A^2$, as in the conformal term.
 Since the dependence on $Q_0$ in Eqs.(\ref{F2Nnew}), (\ref{G2}) and (\ref{eta})  is in the combination $Q_0^2/QQ^\prime$, the modification $Q^\prime_A = \lambda_A  Q^\prime$  can be reabsorbed in the same $Q^2$ rescaling also in the confinement term,  provided that
the confinement length in the nuclear environment scales in the same way: 
\be
Q_0^2 \rightarrow Q_0^2/\lambda_A^2 \,\,\, . \label{Q0resc}
\ee
The origin of the rescaling  (\ref{Qresc}) and (\ref{Q0resc}) in the AdS/CFT framework comes from the identification of  the bulk coordinate with the energy scale of  the dual theory: 
from the form of the $AdS$ metric in Poincar\'e coordinates, a coordinate rescaling $x_\mu \rightarrow \lambda x_\mu$ on the boundary corresponds to $z \rightarrow \lambda z$ in the bulk.
In nuclei, due to the nucleon overlap,  the average distance among quarks and gluons decreases and the color neutralization infrared (confinement) scale increases. 
These modifications in the boundary correspond in the bulk, respectively,  to $z^\prime \rightarrow z^\prime/\lambda$ and $z_0 \rightarrow \lambda z_0$: these are  the prescription (\ref{Qresc}) and (\ref{Q0resc}) used to describe the nuclear effects by redefining the momenta.

In our phenomenological analysis,  the following expression of  the structure function $F_2^A$  (per nucleon) in the nucleus $A$ will be used:
 \be
F_{2}^A(x,Q^2)= F_{2cl}^N\left(x,\frac{Q^2}{\lambda_A^2},Q'\right) + F_{2ct}^N\left(x, \frac{Q^2}{\lambda_A^2},Q',\frac{Q_{0}^2}{\lambda_A^2}\right)\,\,\, . \label{eq:F2A}
\ee
This formula involves the parameter $\lambda_A$, specific of the various nuclei,  to be fitted from data;  moreover, one has to include the proton-neutron difference, discussed below.

\subsection{Deuterium structure function}

\begin{figure}[t]
\includegraphics[width = 0.4\textwidth]{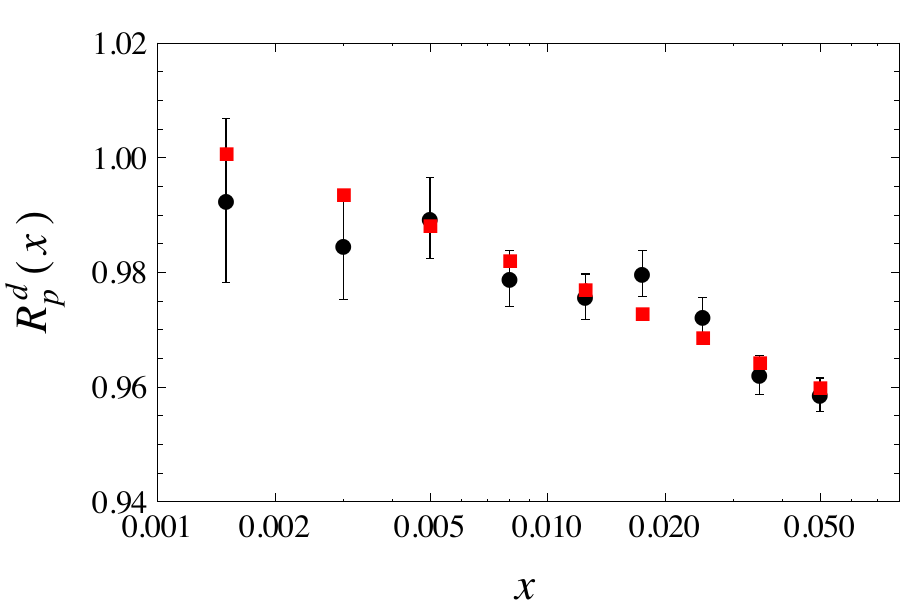}
\caption{Comparison between the  measurements of the ratio of deuterium and proton structure functions $F_2^D/F_2^p$ (black points) \cite{NMC4}  and the expression  obtained by Eqs.~(\ref{deut}), (\ref{deut1}) and (\ref{deut2}) (red squares). In the theoretical formula,  the experimental average $Q^{2}$ for given $x$  is used: the $Q^2$ values (in GeV$^2$), from the first to the last bin in $x$, vary in the range $[0.37-5.8]$. The   $\chi ^{2}$ of the fit is  $\chi ^{2}/d.o.f.=0.85$.}\label{figdp}
\end{figure}

\begin{figure}[b!]
\includegraphics[width=0.4\textwidth]{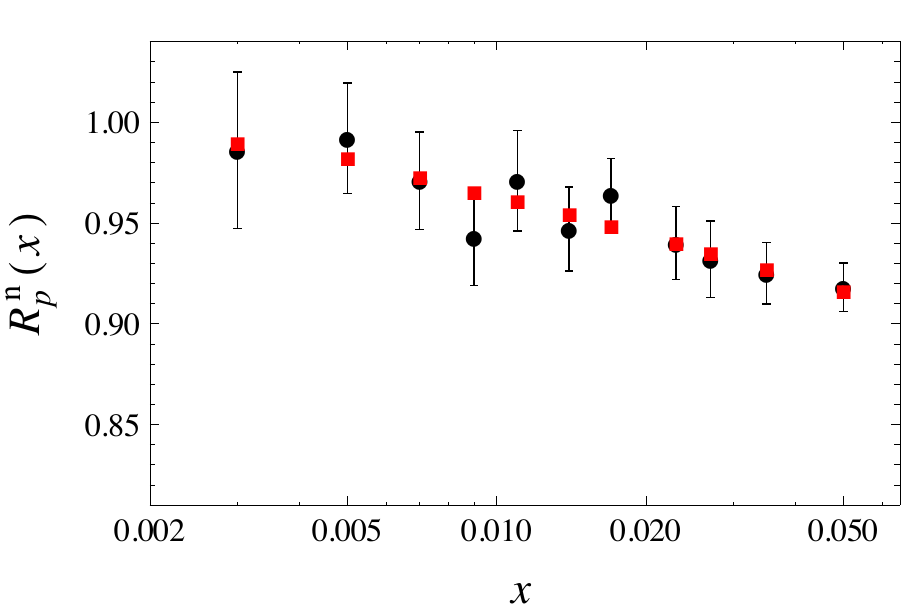}
\caption{Experimental measurements of the ratio  $F_2^n/F_2^p$ (black points) \cite{NMC5}  compared to the  ratio of the   neutron and proton structure function  evaluated by Eq.~(\ref{neut})  (red squares).
In the theoretical expression,   the experimental average $Q^{2}$ for given $x$ is used: from the first to the last bin in $x$, the experimental average $Q^2$  (in GeV$^2$)  varies in the range $[0.4-2.6]$. The   $\chi ^{2}$ of the fit is $\chi ^{2}/d.o.f.=0.23$.}\label{fignp}
\end{figure}

Accounting for  the isospin effects is required in the analysis of  nuclear DIS  data.  We implement such effects  using the neutron  $Q_{0n}$ and $Q^{\prime}_{n}$ scales,  and representing  
the structure function $F_2^D$ (per nucleon) in  deuterium  as
\be
F_2^D = \frac{1}{2} \left[ F_2^{pD}+F_2^{nD}\right] \,\,\, , \label{deut}
\ee
where
\bea
F_2^{pD} &=&  F_2^p\left(x,\frac{Q^2}{\lambda^2_D},Q',\frac{Q_0}{\lambda^2_D}\right)  \,\,\,\, , \label{deut1}\\
F_2^{nD}&=&F_2^n\left(x,\frac{Q^2}{\lambda^2_D},Q'_n,\frac{Q_{0n}}{\lambda^2_D}\right) \,\,\, .\label{deut2}
\eea
Since  deuterium is a weakly bound system,  nuclear effects are small,  and  one expects $\lambda_D \simeq 1$.  Indeed,  a  best fit to data of the expression (\ref{deut}),
shown in Fig.~\ref{figdp},  is  obtained for $\lambda_D=1.011$,  with $Q_{0n}=0.192713$ and $Q^\prime_n=0.177866$.  Using these values of $Q^\prime_n$ and $Q_{0n}$
together with the corresponding parameters  for the proton: $Q_{0p}=0.201613$ and $Q^\prime_p=0.4333$ \cite{Brower:2010wf},
the neutron/proton ratio is  determined, and can be favourably compared to data in Fig.~\ref{figdp} and in Fig.~\ref{fignp}. As expected, the proton and neutron confinement scales $Q_0$ nearly coincide.

\subsection{Heavy nuclei}\label{sec:nuclei}
Before analyzing the nuclear DIS data,
it is worth pointing out that nuclear modifications of the structure functions for heavy nuclei have important phenomenological consequences.
A prime example is the identification of the experimental signatures  of the formation of a possible new state of matter in relativistic heavy ion collisions, at the CERN Large Hadron Collider (LHC) and at the Brookhaven RHIC.  The identification requires a detailed control of the background processes. In the investigation of a possible new state of matter,
the so-called "hard-probes" are crucial, i.e. the dynamical processes originating from  hard-parton scattering. 
The experimental analyses are focused on the differences  in the same phenomenon (jet production, $J/\psi$ suppression, etc.) observed in nucleus-nucleus collisions with respect to proton-proton and proton-nucleus scattering,  where the obtained energy density is  not enough to produce the transition to the new phase.
Since the hard-parton scattering involves the parton distribution functions (pdfs), statements on the experimental signature of the new state of matter using hard-probes crucially depend on the control   on the  modifications of structure functions induced by the ordinary nuclear dynamics \cite{eskola1,salgado1,yellow}. 

\begin{figure}[t]
\includegraphics[width = 0.34\textwidth]{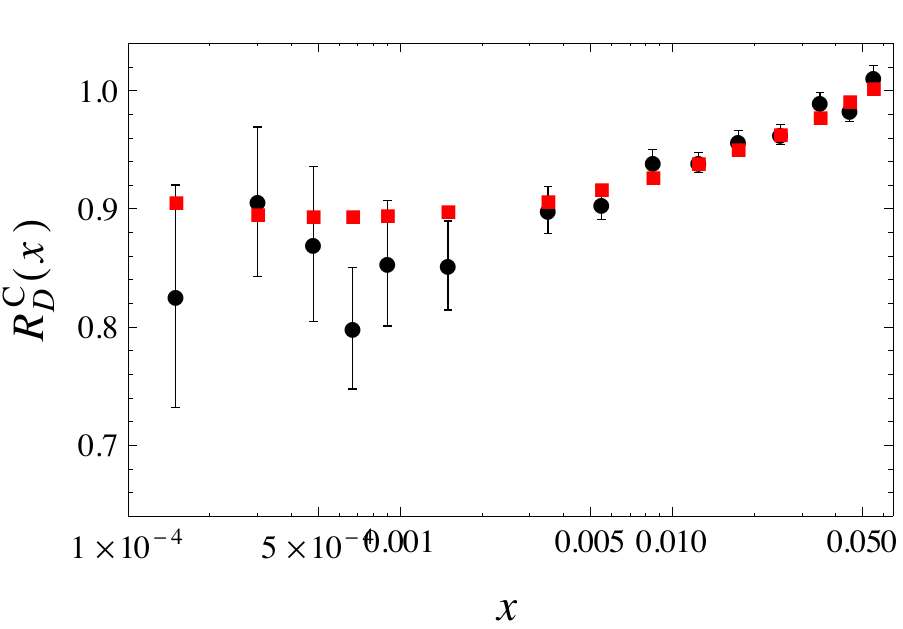}\\
\includegraphics[width = 0.34\textwidth]{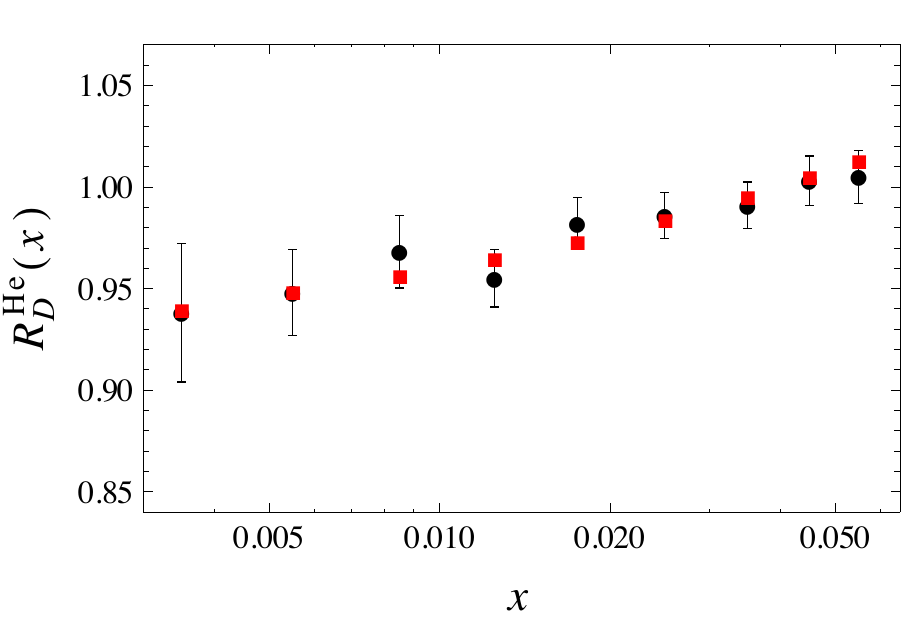} \\
\includegraphics[width = 0.34\textwidth]{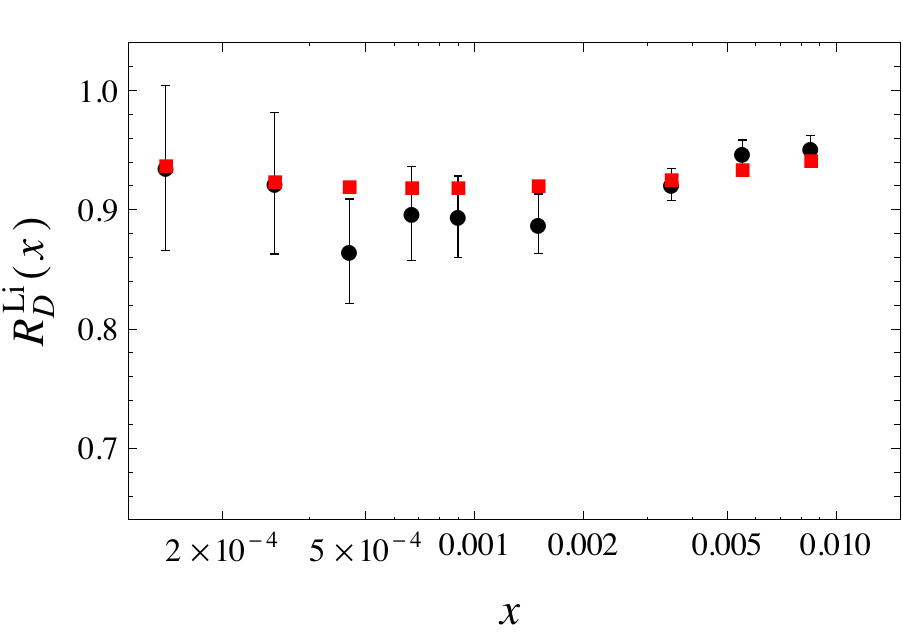}\\
\includegraphics[width = 0.34\textwidth]{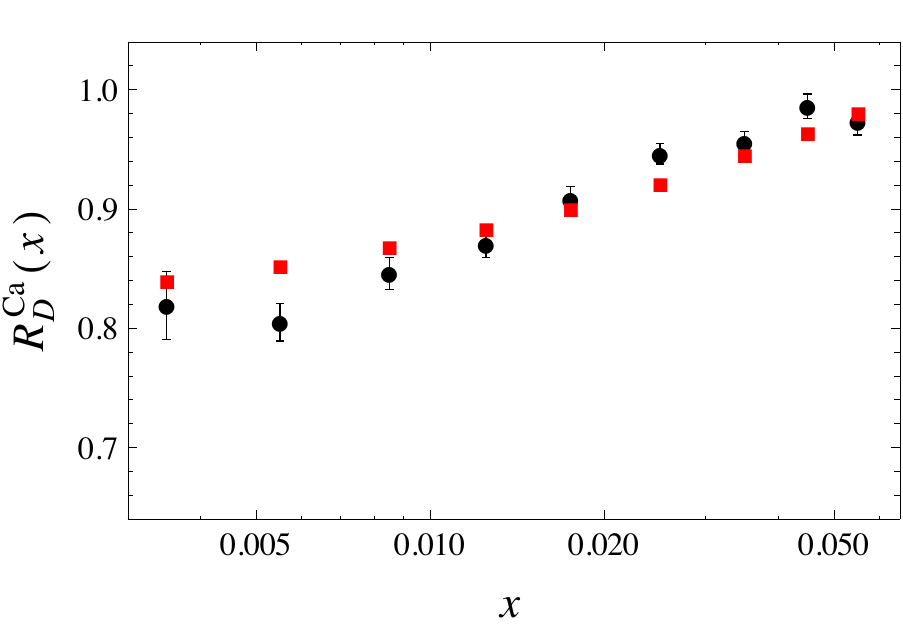}
\caption{Ratio $F_2^A/F_2^D$ for various nuclei. The black points correspond to the experimental measurements with the data sets in Table~\ref{tab:points},  the red boxes to the  holographic formulae with parameters $\lambda_A$ in Table~\ref{tab:HeD}. The isospin breaking effect has been taken into account.  From top-down, the panels correspond to:  C/D, He/D, Li/D, Ca/D.  The  $\chi^2_{d.o.f}$ of the fit of the structure functions is in Table~\ref{tab:points}. }\label{fig:res1}
\end{figure}

\begin{figure}[h]
\includegraphics[width = 0.34\textwidth]{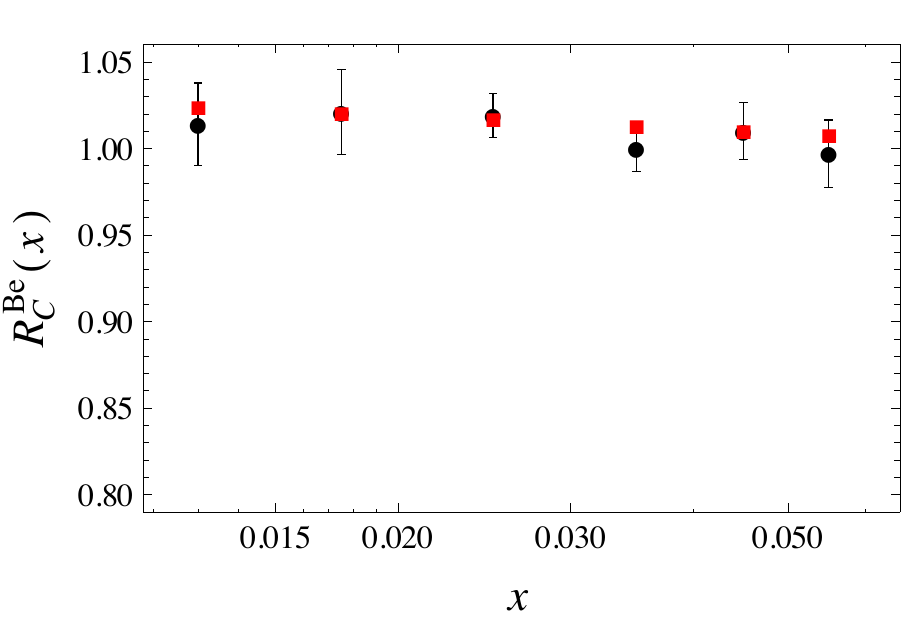}\\
\includegraphics[width = 0.34\textwidth]{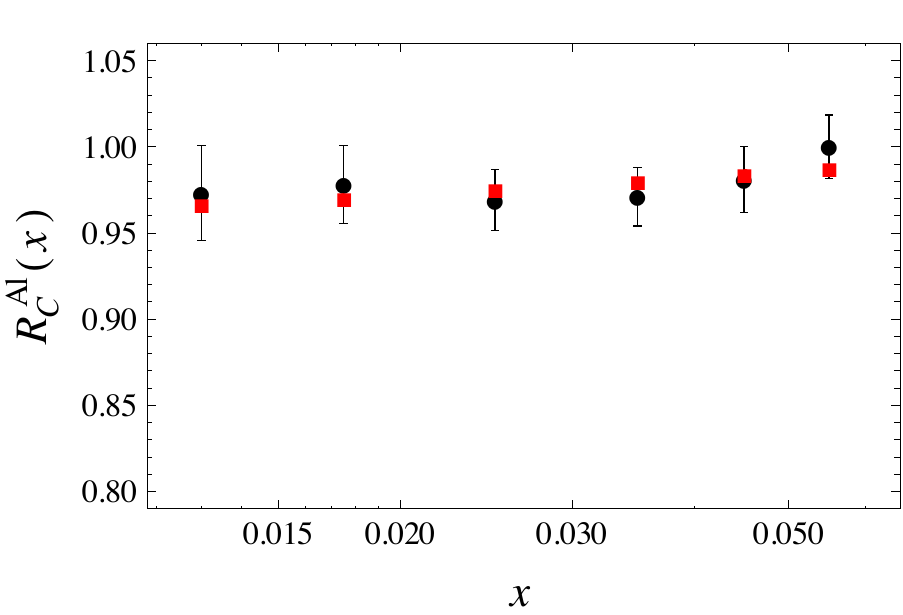}\\
\includegraphics[width = 0.34\textwidth]{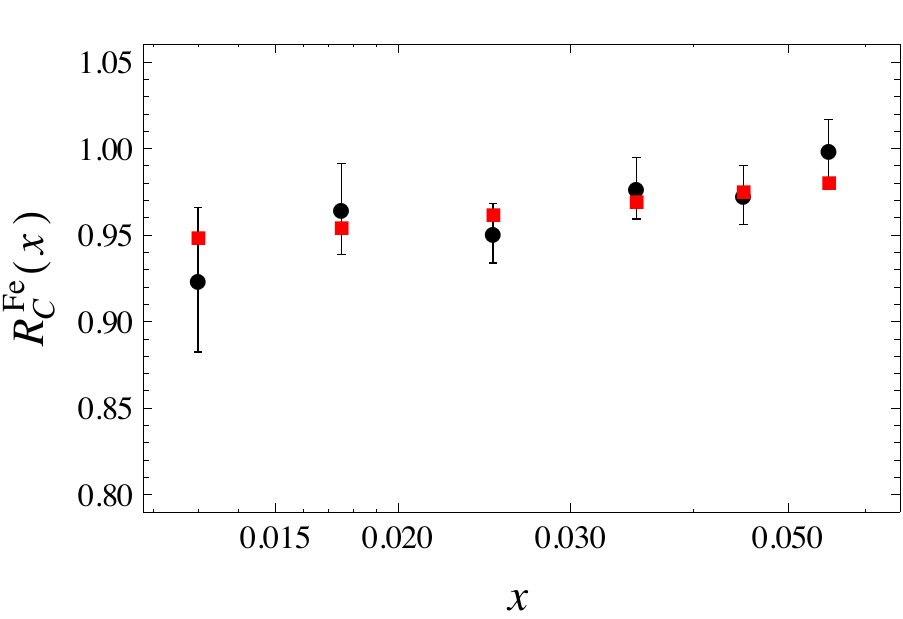}\\
\includegraphics[width = 0.34\textwidth]{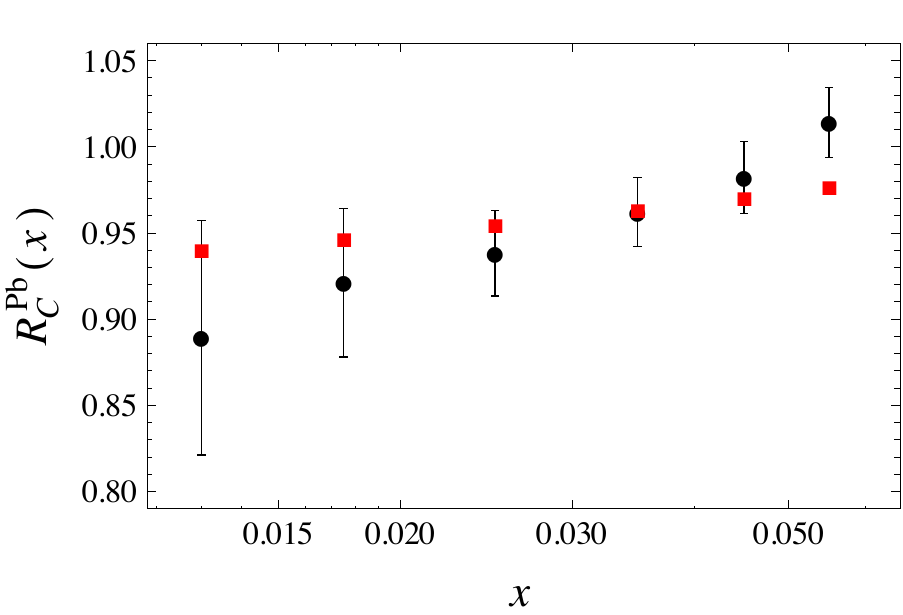}
\caption{Ratio $F_2^A/F_2^C$ for various nuclei. The black points correspond to the experimental measurements with the data sets in Table~\ref{tab:points},  the red boxes to the  holographic formulae with parameters $\lambda_A$ in Table~\ref{tab:HeD}. The isospin breaking effect has been taken into account.  From top-down, the panels correspond to:
Be/C, Al/C, Fe/C, Pb/C. The  $\chi^2_{d.o.f}$ are  in Table~\ref{tab:points}.}\label{fig:res2}
\end{figure}

Coming to the analysis of nuclear DIS data, 
the holographic expression of $F_2^A$ for a nucleus with charge $Z$ can be written as
\bea
F_2^A(x,Q^2)&=&\left(\frac{Z}{A}\right) F_2^p\left(x,\frac{Q^2}{\lambda_A^2},Q^\prime,\frac{Q_0}{\lambda_A^2}\right) \,\,\,\,\,\, \,\,\, \nonumber \\
&+&\left(1-\frac{Z}{A}\right) F_2^n\left(x,\frac{Q^2}{\lambda_A^2},Q^\prime_n,\frac{Q_0}{\lambda_A^2}\right),
\eea
with  the proton and neutron structure functions  in Eqs.(\ref{F2p})  and (\ref{neut}), and  the scaling parameter $\lambda_A$  accounting for the nuclear modification.
 For different nuclei,  the ratio $R_A=F_2^A/F_2^D$ can be analytically evaluated at small $Q^2$ and small $x$,  in a regime  where the perturbative approach cannot be applied. 
The results can be compared to the experimental data, using  the  data sets  in Table \ref{tab:points}  for the various nuclei, together with the values of $\lambda_A$    in Table~\ref{tab:HeD}. The  comparison is shown in Figs.~\ref{fig:res1} and \ref{fig:res2}.  Considering
the $\chi^2/d.o.f.$  reported in Table~\ref{tab:points}  for each nucleus,  the agreement of the theoretical formula with data is remarkable, and the $x-$dependence exhibited by data is closely followed by the theoretical results.

It is interesting to comment on the isospin breaking effects, since fits of the nuclear structure functions could  also be done neglecting the proton-neutron difference. The inclusion of  the isospin effect improves the accuracy of the fits, as one can infer from the various $\chi^2_{d.o.f}$  in Table~\ref{tab:points}; the only exceptions are Be and Fe, where 
$\chi^2_{d.o.f}$ remain essentially unchanged if the isospin breaking is considered.

\begin{table}[h]
	\centering
	\begin{tabular}{c c c c c c}
\hline
	  nucleus\,\, &n. points\,\,&$\chi^2_{d.o.f}$\,\,&n. points\,\,&$\chi^2_{d.o.f}$\,\,&range of $\left\langle Q^2\right\rangle$\\
	\hline    
He	&9&	$1.09$	&   9 &$0.24 $ & $[0.77-6.3]$\\
Li	&9&	$0.93$	&   9 &$0.79 $ & $[0.03-1.4]$\\
Be	&6&	$0.21$	&   6 &$0.30 $ & $[3.4-11.4]$\\
C	&9&	$1.61$	&  15 &$0.89 $ & $[0.03-6.4]$\\
Al	&6&	$0.23$	&   6 &$0.21 $ & $[3.4-11.6]$\\
Ca	&9&	$8.0$	&   9 &$3.87 $ & $[0.6-6.8]$\\
Fe	&6&	$0.41$	&   6 &$0.42 $ & $[3.4-11.8]$\\
Pb 	&6&	$1.11$	&   6 &$0.93 $ & $[3.4-11.6]$\\
\hline
  \end{tabular}
	\caption{Experimental data sets  \cite{NMC} and $\chi^2_{d.o.f}$ of the fit of the structure function $F_2^A$  for each nucleus. The third column reports the $\chi^2_{d.o.f}$  of  fits without isospin breaking,  the fourth and fifth columns correspond to fits  with  the isospin breaking effect included. In the last column,  the experimental average $Q^2$ ranges (in GeV$^2$) for the various cases are indicated, from the first to the last bin of the Bjorken $x$. }\label{tab:points}
	\end{table}

\subsection{$x$-rescaling}

\begin{figure}[t]
\includegraphics[width=0.4\textwidth]{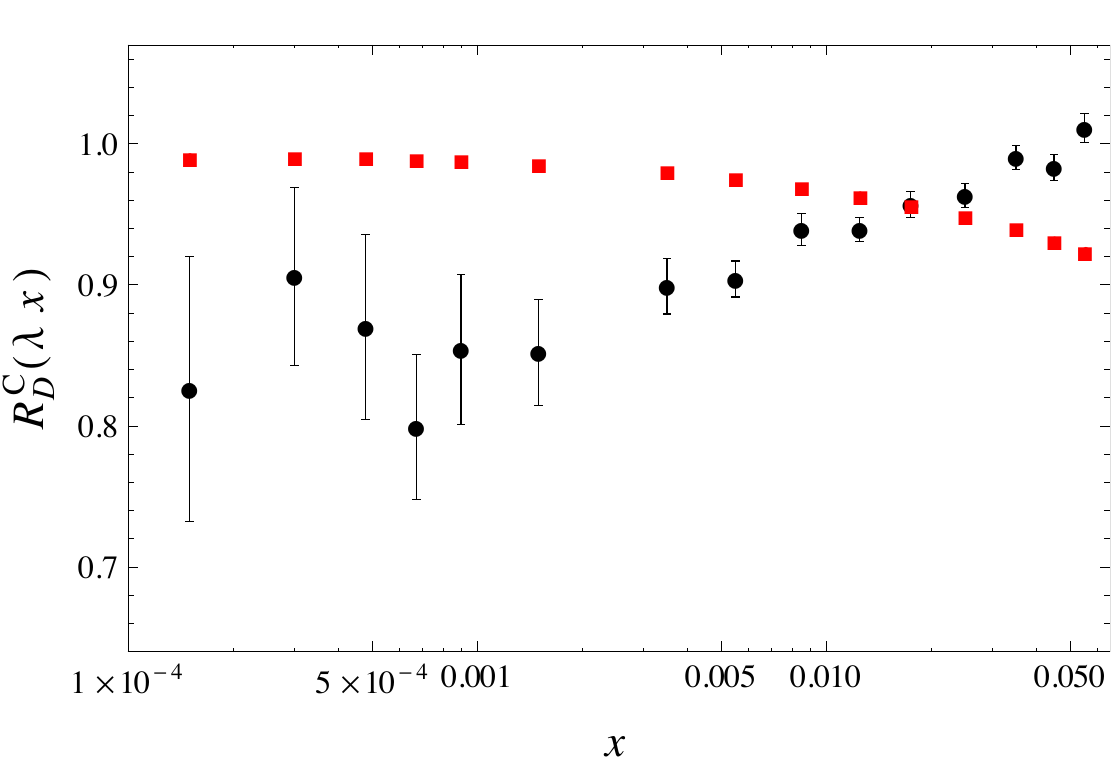}
\caption{Comparison of experimental data for the ratio $F_2^C/F_2^D$ (black points) with the result obtained by  $x$-rescaling in the holographic expression of the structure function (red boxes).}\label{fig2}
\end{figure}

We have shown that, in the holographic approach, the nucleon structure function $F_2^A$ at low-$x$ in a nuclear environment can be  obtained rescaling the effective lengths appearing  in the nucleon wave function in  nuclei.
It is interesting to notice that, using the local approximation (\ref{localP24}) and (\ref{localP13}), one has the combination $\tau = \log{(\rho Q / 2 x Q^\prime)}$.  Therefore, 
the rescaling $Q^\prime_A = \lambda_A  Q^\prime$ could be reabsorbed not in the $Q^2$ rescaling, but rather in  $x \rightarrow \lambda_A x$. However, due to the $Q^2$ dependence of $F_2$ in Eq.(\ref{F2Nconf1}), the $x$-rescaling is not  equivalent to the  $Q^2$ rescaling,   and consequently $F_2^A(x,Q^2/\lambda^2) \neq F_2^A(\lambda x, Q^2)$.
One can wonder if the $x$-rescaling is  in agreement with data: looking at  Fig.~\ref{fig2} we conclude that this is not the case,  not surprisingly,  since the $x-$rescaling has been proposed as a possible explanation of the EMC effect at large-$x$.

\subsection{Remarks}\label{sec:remarks}

As we have discussed, in the  holographic formula the nuclear effects in the DIS structure functions can be described by a  $Q^2$-rescaling,   corresponding to a modification of the confinement length for a bound nucleon.
Other different methods  produce similar results. An example is the QCD dipole model \cite{dipole1,dipole2}, where the structure functions are determined
considering  a virtual photon $\gamma^*$ splitting in a quark-antiquark dipole which interacts with the target $T$. 
Encoding the energy and target size dependence of the dipole-target cross section
 $\sigma^{\gamma^* T}$  in the saturation scale $Q_{S,T}(x)$ 
\cite{albacete}, $\sigma^{\gamma^* T}$ turns out to
depend only on the ratio $\tau^2_T = Q^2/Q^2_{S,T}(x)$. This implies a   geometric scaling between the nucleus and the nucleon cross sections \cite{albacete}:
\be
\frac{\sigma^{\gamma^* A}(\tau_A)}{\pi R_A^2}=\frac{\sigma^{\gamma^* N}(\tau_N)}{\pi R_N^2} \,\,\, ,
\ee
with radii $R_{N,A}$ and
\be
\tau^2_A=\tau^2_N\left(\frac{\pi R_A^2}{A \pi R_N^2}\right)^{1/\delta}\,\,\,\ .
\ee
The consequence is
\be
Q^2_{S,A} = Q^2_{S,N}\left(\frac{A \pi R_N^2}{\pi R_A^2}\right)^{1/\delta}  \,\,\,\ .
\ee
Since the cross section  only depends on  $Q^2/Q^2_{S,T}(x)$,  the replacement $Q^2_{S,N} \rightarrow Q^2_{S,A}$ corresponds to  rescaling 
\be
Q^2 \rightarrow Q^2/\lambda^2_{A,dip} \,\,\, , 
\ee
with
\be
\lambda_{A,dip} = \left(\frac{A \pi R_N^2}{\pi R_A^2}\right)^{1/2\delta}.
\ee
In the dipole model  low-$x$ nuclear data  are reproduced  for $R_A = (1.12 A^{1/3}-0.86 A^{-1/3})$ fm, $\pi R^2_N = 1.55$ fm$^2$, and $\delta = 0.79$ \cite{albacete}.
\begin{table}[h]
	\centering
	\begin{tabular}{c c c }
\hline
	  nucleus \,\,\,&$\lambda_A$(holography)\,\,&$\lambda_{A,dip}$\cite{albacete}\\
	\hline    
Li	&	$1.843$	&   $1.130$ \\
Be	&	$1.764$	&   $1.140$ \\
C	&	$1.775$	&   $1.160$ \\
Al	&	$1.972$	&   $1.264$ \\
Ca	&	$2.006$	&   $1.338$ \\
Fe	&	$2.090$	&   $1.413$ \\
Pb 	&	$2.286$	&   $1.780$ \\
\hline
  \end{tabular}
	\caption{Rescaling parameter $\lambda_A$ obtained  using  the holographic expression for $F_2^A$ and taking into account the isospin breaking.  The values  in the last column are obtained within  the QCD dipole model 
\cite{albacete}. }\label{tab:HeD}
	\end{table}

In Table \ref{tab:HeD} we compare the rescaling parameters $\lambda_A$ obtained in  the holographic   and in the QCD dipole model.  Regardless of  the difference between the two theoretical approaches,  the rescaling parameters  differ by less than $30-35\%$; however, the deviation is larger than in the case where the isospin breaking is neglected \cite{noi}.

\section{Nuclear modification of the longitudinal structure function}\label{sec:FL}
The experimental determination of the structure function per nucleon in a nucleus is usually done by cross section data, assuming  a minor nuclear effect on the longitudinal structure function $F_L = F_2 - 2xF_1$,  hence using the value of the free nucleon   $F_L^N$.
This procedure has to be checked, because it introduces an uncertainty in the evaluation of the nuclear structure functions which, in turn, implies an uncertainty in the determination of the modified pdfs.

\begin{figure}[h]
\includegraphics[width = 0.34\textwidth]{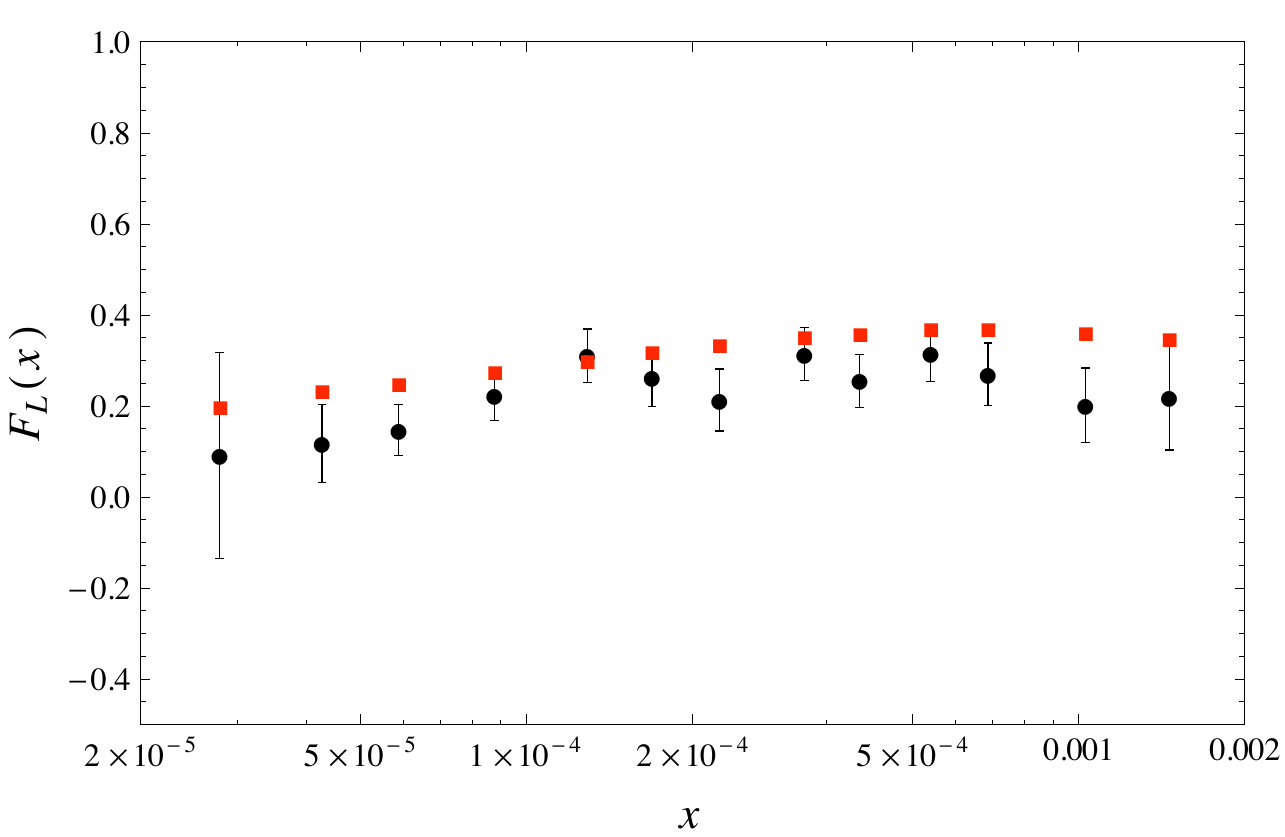}
\caption{Comparison with experimental data (black dots) \cite{h1} of the  longitudinal structure function of  the proton evaluated using the holographic formulae (red squares).  The experimental $\langle Q^2\rangle$, from the first  to the last bin in $x$,  varies in the range $[1.5-45]$ GeV$^2$. The $\chi^2_{d.o.f}$ is  $\chi^2_{d.o.f}\simeq 1.1$.}\label{fig:Flproton}
\end{figure}

\begin{figure}[b]
\includegraphics[width = 0.34\textwidth]{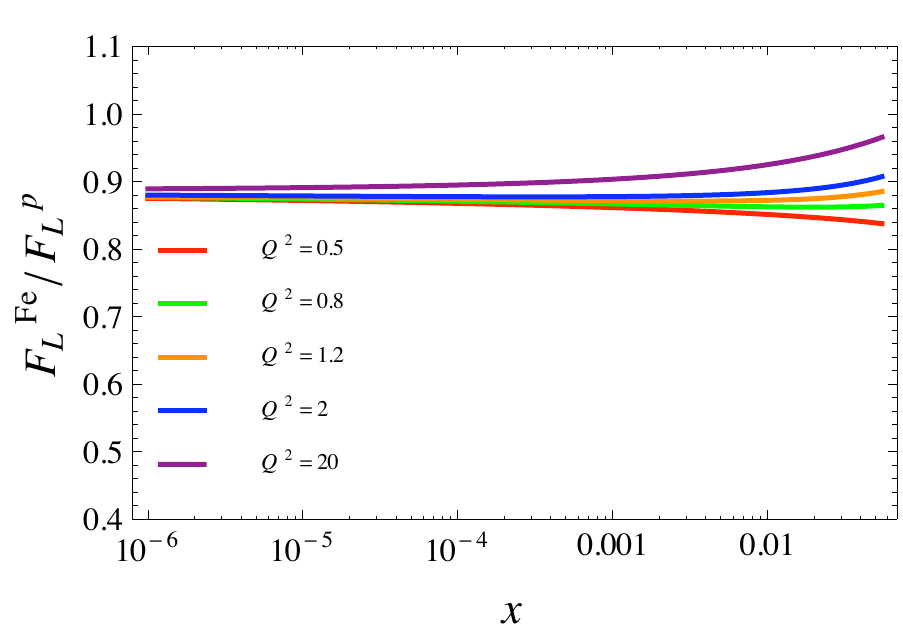}\\
\includegraphics[width = 0.34\textwidth]{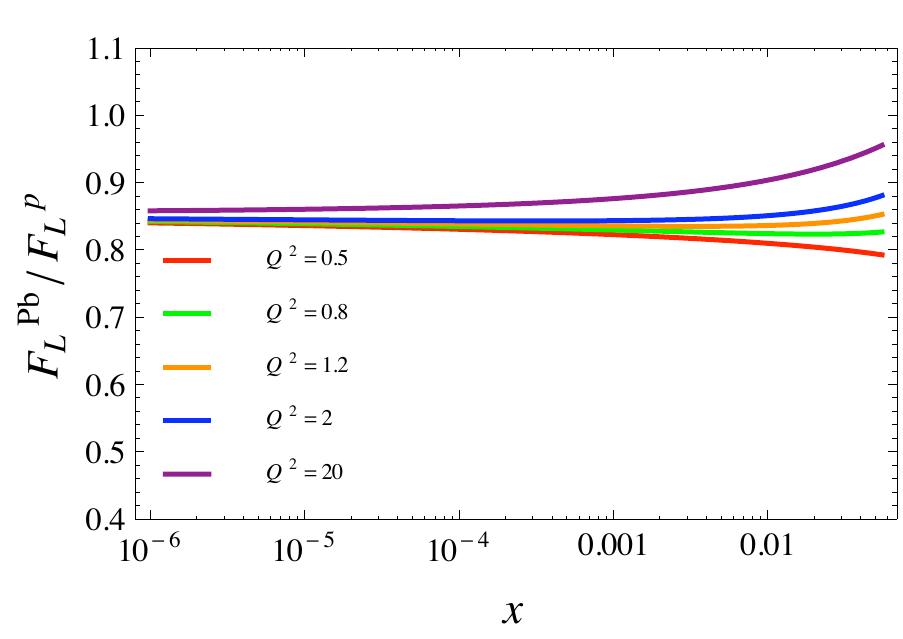} 
\caption{ Ratio between the longitudinal nuclear structure function per nucleon  and for free nucleon. From top-down, the panels correspond to  Fe and Pb. }\label{fig:FL1}
\end{figure}

A holographic expression for the longitudinal structure function $F_L$ can obtained from Eqs.~(\ref{F2Nconf}) and (\ref{F2Nnew}), using the local approximation for $P_{24}$, and for
$P_{13}$  
\be
P_{13}(z,Q^2)|_{F_L} = \frac{1}{z} (Qz)^2  K_0^2(Qz)  \,\,\ . \label{P13L}
\ee
For the proton,  the comparison with the experimental data \cite{h1} is shown in Fig.~\ref{fig:Flproton}.
For the  nuclear case,  using the values of the parameters determined above, we obtain for the ratio $F_L^A/F_L^p$  the results  in Fig.~\ref{fig:FL1}.

\begin{figure}[t]
\includegraphics[width = 0.34\textwidth]{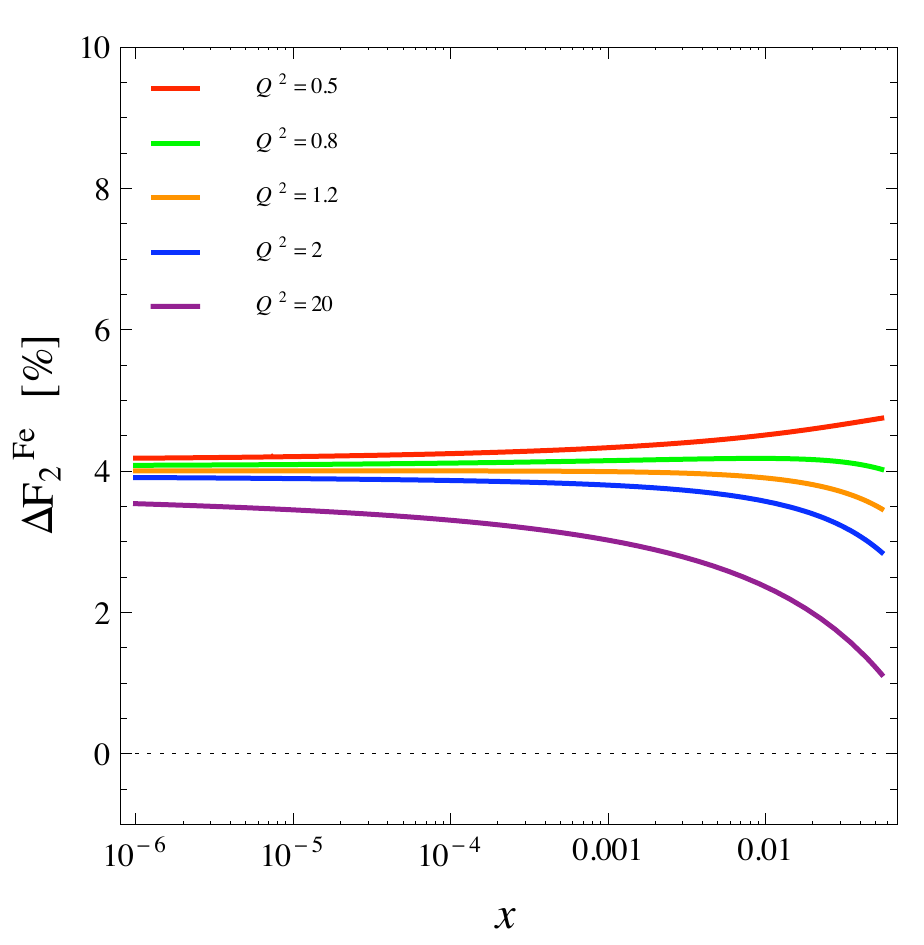}\\
\includegraphics[width = 0.34\textwidth]{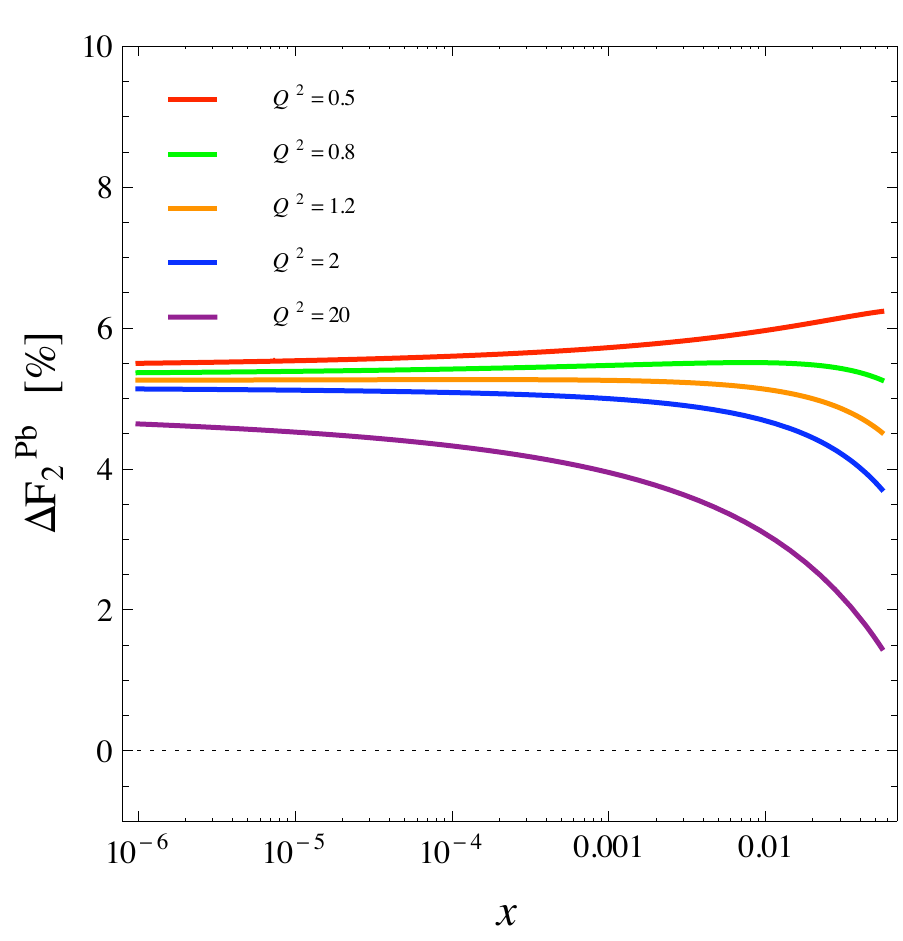} 
\caption{Maximum uncertainty ($y=1$) in the experimental determination of the structure function $F_2^A$  due to the absence of nuclear effects in $F_L$.  The top panel corresponds to  Fe, the bottom one to  Pb.}\label{fig:delta}
\end{figure}

In order to evaluate the uncertainty in the extraction of the nuclear structure functions, we recall that
the structure function  is experimentally determined by  data on the reduced cross section  $\sigma_r$: 
\be
\sigma_r = F_2 \left[1- f(y) \frac{F_L}{F_2} \right]\,\,\ ,
\ee
where 
\be
f(y)= \frac{y^2}{1+(1-y)^2} \,\,\ .
\ee
 Let us call $\hat F_2^A$ the structure function per nucleon obtained by the relation
\be
\sigma_r = \hat F_2^A - f(y) F_L^N   \,\, , 
\ee
i.e.,  using the longitudinal structure function  of the free nucleon,  without nuclear effects.
$\hat F_2^A$ is an approximation of  $F_2^A$ which should be determined by the relation
\be
\sigma_r =  F_2^A - f(y) F_L^A  .
\ee
By the  expression of $F_2^A$ in Eq.~(\ref{eq:F2A}),  and using  the previous equations, one can  evaluate the uncertainty on $F_2^A$:
\be
\Delta F_2^A= \frac{\hat F_2^A - F_2^A}{\hat F_2^A} = 1 -\frac{F_2^A}{F_2^A+f(y)(F_L^N-F_L^A)} \,\,\, .
\ee
As shown in Fig.~\ref{fig:delta},  the maximum uncertainty (corresponding to $y=1$) in the extraction of $F_2^A$ is of few percent also in the region of very low $x$ and $Q^2$.
This is consistent with the results in  Ref.~\cite{Armesto:2010tg}, where the longitudinal structure function in nuclear DIS at small $x$ and  $Q^2\ge 4$ GeV$^2$ is discussed in the framework of universal parton densities obtained in DGLAP analysis at next-leading-order (NLO), with the conclusion that the uncertainty in  $F_2^A$ is less than $10\%$.\\

\section{Conclusions and perspectives}\label{sec:conclusions}
 
A description of nuclear shadowing, i.e. the distortion at low-$x$ of the nuclear DIS structure functions,  can be obtained  by a rescaling the  virtual photon momentum  $Q^2$,  and   this  modification naturally emerges  in a holographic  approach.
Experimental data for electroproduction are theoretically reproduced, hence  the AdS/CFT formulation  captures  the relevant  dynamics  to describe the nuclear DIS effects. 

The next step of the study would be  the analysis of the experimental results for DIS neutrino  scattering on nuclear target, an  interesting issue due to the large  theoretical uncertainties in current calculations of neutrino  cross section at  high energy and very low $x$  \cite{kuroda}.
Universality of nuclear effects in DIS has been recently shown \cite{neutrino1} by the analysis of neutrino data which takes into account the different normalizations of independent experiments: the nuclear modifications are found to be the same as in electroproduction. 
A  calculation in the holographic framework would require the solution of the equation of motion for charged currents in the bulk,  to obtain an expression analogous to (\ref{P13}): this   analysis  deserves a dedicated study. For the time being,  simple arguments are encouraging.  Indeed,
 for a correct normalization procedure and to facilitate the data comparison with theory, in Ref. \cite{neutrino1}  the ratio between neutrino data on nuclear target and the theoretical proton cross-section (i.e. without nuclear effects) are considered, instead of the absolute experimental cross section.
The average value of this ratio, $R^\nu_A$, in the small-$x$ bins,  turns out to be  
 $R^\nu_A \simeq 0.94 \pm 0.09$ for $x =0.015$, $R^\nu_A \simeq 1 \pm 0.08$ for $x =0.045$ and $R^\nu_A \simeq 1.03 \pm 0.05$ for $x=0.08$ \cite{neutrino1}. 
Neglecting the contribution of the structure function $xF_3$ to the cross section, which should be small in the considered kinematical region, a  comparison can be done between  $R^\nu_A$ and the ratio $F_2^{Fe}/F_2^p$ evaluated in the holographic approach for the corresponding average values of $x$ and $Q^2$.
One obtains:  $F_2^{Fe}/F_2^p \simeq 0.88,0.93,0.97$ for $x = 0.015,0.045,0.08$, respectively,  consistent with the corresponding $R^\nu_A$.
The  conclusion is that the approach based on the holographic method  is also promising  for other  analyses, for instance neutrino scattering, and that the method can be applied to  small  $Q^2$ values, confirming the complementarity  of the AdS/CFT  inspired techniques  with  the perturbative calculations.

\acknowledgments{Paolo Castorina acknowledges the CERN TH Unit for hospitality. }


\begin{thebibliography}{99}


\bibitem{arneodo} 
For a review see: M.~Arneodo,
Phys.\ Rept.\  {\bf 240}, 301 (1994).

\bibitem{Frankfurt:2011cs} 
  L.~Frankfurt, V.~Guzey and M.~Strikman,
  Phys.\ Rept.\  {\bf 512}, 255 (2012).
 

\bibitem{aku} S.~V.~Akulinichev et al., Phys.\ Lett.\ B\ {\bf 158}, 485 (1985);
Phys.\ Rev.\ Lett.\  {\bf 55},  2239 (1985).

\bibitem{close}
  F.~E.~Close, R.~G.~Roberts and G.~G.~Ross,
  Phys.\ Lett.\ B {\bf 129}, 346 (1983);
  F.~E.~Close, R.~L.~Jaffe, R.~G.~Roberts and G.~G.~Ross,
  Phys.\ Rev.\ D {\bf 31}, 1004 (1985);
  R.~L.~Jaffe,
  Phys.\ Rev.\ Lett.\  {\bf 50}, 228 (1983).

\bibitem{close2} 
  R.~L.~Jaffe, F.~E.~Close, R.~G.~Roberts and G.~G.~Ross,
  Phys.\ Lett.\ B {\bf 134}, 449 (1984).

\bibitem{close4}  
  F.~E.~Close, R.~G.~Roberts and G.~G.~Ross,
  Phys.\ Lett.\ B {\bf 168}, 400 (1986);
  R.~P.~Bickerstaff and G.~A.~Miller,
  Phys.\ Lett.\ B {\bf 168}, 409 (1986).

\bibitem{castorina}   
P.~Castorina and A.~Donnachie,
  Phys.\ Lett.\ B {\bf 215}, 589 (1988),
  Z.\ Phys.\ C {\bf 45}, 141 (1989).

\bibitem{pov}   
B.~Povh and J.~Hufner,
Phys.\ Rev.\ Lett.\  {\bf 58}, 1612 (1987).

\bibitem{adscft} 
J.~M.~Maldacena,
Adv.\ Theor.\ Math.\ Phys.\ \ {\bf 2}, 231 (1998) 
[Int.\ J.\ Theor.\ Phys.\ \ {\bf 38}, 1113 (1999)];
%
S.~S.~Gubser, I.~R.~Klebanov and A.~M.~Polyakov,
Phys.\ Lett.\ B\ {\bf 428}, 105 (1998);
%
E.~Witten,
Adv.\ Theor.\ Math.\ Phys.\ \ {\bf 2}, 253 (1998).
 
\bibitem{finite} 
  See, e.g., 
 G.~F.~de Teramond and S.~J.~Brodsky,
  Phys.\ Rev.\ Lett.\  {\bf 94}, 201601 (2005);
J.~Erlich, E.~Katz, D.~T.~Son and M.~A.~Stephanov,
  Phys.\ Rev.\ Lett.\  {\bf 95}, 261602 (2005);
 O.~Aharony, J.~Sonnenschein and S.~Yankielowicz,
  Annals Phys.\  {\bf 322}, 1420 (2007);
J.~Casalderrey-Solana, H.~Liu, D.~Mateos, K.~Rajagopal and U.~A.~Wiedemann,
  arXiv:1101.0618 [hep-th].

\bibitem{Polchinski} 
J.~Polchinski and M.~J.~Strassler,
Phys.\ Rev.\ Lett.\  {\bf 88}, 031601 (2002),
JHEP {\bf 0305}, 012 (2003).

\bibitem{other-dis}
C.~A.~Ballon Bayona, H.~Boschi-Filho and N.~R.~F.~Braga,
  JHEP {\bf 0803}, 064 (2008), 
  JHEP {\bf 0810}, 088 (2008);
    L.~Cornalba and M.~S.~Costa,
  Phys.\ Rev.\ D {\bf 78}, 096010 (2008);
    L.~Cornalba, M.~S.~Costa and J.~Penedones,
  JHEP {\bf 1003}, 133 (2010),
  Phys.\ Rev.\ Lett.\  {\bf 105}, 072003 (2010);
    Y.~Hatta, E.~Iancu and A.~H.~Mueller,
  JHEP {\bf 0801}, 026 (2008),
  JHEP {\bf 0801}, 063 (2008);
    J.~L.~Albacete, Y.~V.~Kovchegov and A.~Taliotis,
  JHEP {\bf 0807}, 074 (2008);
    B.~Pire, C.~Roiesnel, L.~Szymanowski and S.~Wallon,
  Phys.\ Lett.\ B {\bf 670}, 84 (2008);
  Y.~V.~Kovchegov, Z.~Lu and A.~H.~Rezaeian,
  Phys.\ Rev.\ D {\bf 80}, 074023 (2009);
    A.~H.~Mueller, A.~I.~Shoshi and B.~-W.~Xiao,
  Nucl.\ Phys.\ A {\bf 822} (2009) 20;
    E.~Avsar, E.~Iancu, L.~McLerran and D.~N.~Triantafyllopoulos,
  JHEP {\bf 0911}, 105 (2009);
    C.~Marquet, B.~-W.~Xiao and F.~Yuan,
  Phys.\ Lett.\ B {\bf 682}, 207 (2009);
    Y.~V.~Kovchegov,
  Phys.\ Rev.\ D {\bf 82}, 054011 (2010);
    A.~Watanabe and K.~Suzuki,
  Phys.\ Rev.\ D {\bf 86}, 035011 (2012).

  \bibitem{Brower:2010wf}
R.~C.~Brower, M.~Djuric, I.~Sarcevic and C.~-I~Tan,
JHEP {\bf 1011}, 051 (2010).

\bibitem{noi}
  L.~Agozzino, P.~Castorina and P.~Colangelo,
  arXiv:1306.5072 [hep-ph], to appear in PRL.

\bibitem{Brower:2006ea}
R.~C.~Brower, J.~Polchinski, M.~J.~Strassler and C.~-I~Tan,
JHEP {\bf 0712}, 005 (2007).

\bibitem{NMC4} 
  M.~Arneodo {\it et al.}  [New Muon. Collaboration],
  Nucl.\ Phys.\ B {\bf 487}, 3 (1997).

\bibitem{NMC5} 
  P.~Amaudruz {\it et al.}  [New Muon Collaboration],
  Nucl.\ Phys.\ B {\bf 371}, 3 (1992).

\bibitem{eskola1}
  K.~J.~Eskola, H.~Paukkunen and C.~A.~Salgado,
  JHEP {\bf 0904}, 065 (2009).

\bibitem{salgado1}   K.~J.~Eskola, V.~J.~Kolhinen, H.~Paukkunen and C.~A.~Salgado,
  JHEP {\bf 0705}, 002 (2007).

\bibitem{yellow}   
A.~Accardi, F.~Arleo, N.~Armesto, R.~Baier, D.~G.~d'Enterria, R.~J.~Fries, O.~Kodolova and I.~P.~Lokhtin {\it et al.},
  hep-ph/0310274.

\bibitem{NMC} 
  P.~Amaudruz {\it et al.}  [New Muon Collaboration],
  Nucl.\ Phys.\ B {\bf 441}, 3 (1995);
  M.~Arneodo {\it et al.}  [New Muon. Collaboration],
  Nucl.\ Phys.\ B {\bf 441}, 12 (1995);
  M.~Arneodo {\it et al.}  [New Muon Collaboration],
  Nucl.\ Phys.\ B {\bf 481}, 3 (1996).

\bibitem{dipole1} 
  N.~N.~Nikolaev and B.~G.~Zakharov,
  Z.\ Phys.\ C {\bf 49}, 607 (1991).

\bibitem{dipole2} 
  A.~H.~Mueller,
  Nucl.\ Phys.\ B {\bf 415}, 373 (1994).

\bibitem{albacete} 
  J.~L.~Albacete, N.~Armesto, J.~G.~Milhano, C.~A.~Salgado and U.~A.~Wiedemann,
  Eur.\ Phys.\ J.\ C {\bf 43}, 353 (2005),
  Phys.\ Rev.\ D {\bf 71}, 014003 (2005).
  
\bibitem{h1} 
  F.~D.~Aaron {\it et al.}  [H1 Collaboration],
  Eur.\ Phys.\ J.\ C {\bf 71}, 1579 (2011).

\bibitem{Armesto:2010tg}
  N.~Armesto, H.~Paukkunen, C.~A.~Salgado and K.~Tywoniuk,
  Phys.\ Lett.\ B {\bf 694}, 38 (2010).

\bibitem{kuroda} 
  M.~Kuroda and D.~Schildknecht,
  Phys.\ Rev.\ D {\bf 88}, 053007 (2013).

\bibitem{neutrino1} 
  H.~Paukkunen and C.~A.~Salgado,
  Phys.\ Rev.\ Lett.\  {\bf 110}, 212301 (2013).

\end{thebibliography}
\end{document}